\documentclass[useAMS,usenatbib]{mn2e}
\usepackage{graphicx}
\usepackage{amsmath}
\usepackage{color}
\usepackage{amssymb}

\newcommand{\Ion}[2]{#1{\,\sc#2}}
\newcommand{\Rwd}{\mbox{$R_{\mathrm{wd}}$}}
\newcommand{\Mwd}{\mbox{$M_{\mathrm{wd}}$}}
\newcommand{\Msun}{\mbox{$\mathrm{M}_{\odot}$}}
\newcommand{\Rsun}{\mbox{$\mathrm{R}_{\odot}$}}

\begin{document}

\title[Composition Of An Extrasolar Planetesimal]{The Composition Of A Disrupted Extrasolar Planetesimal At SDSS\,J0845+2257 (Ton\,345)}
\author[D.J. Wilson et
  al.]{D. J. Wilson$^1$\thanks{d.j.wilson.1@warwick.ac.uk},
  B. T. G{\"a}nsicke$^1$, D. Koester$^2$, O. Toloza$^1$, A. F. Pala$^1$, \newauthor E. Breedt$^1$, S. G. Parsons$^3$  \medskip\\
$^{1}$ Department of Physics, University of Warwick, Coventry CV4 7AL,
UK\\
$^{2}$ Institut f\"ur Theoretische Physik und Astrophysik, University of Kiel,
24098 Kiel, Germany\\
$^4$ Departamento de F\'isica y Astronom\'ia, Universidad de Valpara\'iso,
Avenida Gran Breta\~na 1111, Valpara\'iso 2360102, Chile \\
} \date{Accepted 2015 May 22. Received 2015 May 22; in original form 2015 February 20}
\maketitle

\begin{abstract}
We present a detailed study of the metal-polluted DB white dwarf SDSS\,J0845+2257 (Ton\,345). Using high-resolution {\em HST}/COS and VLT spectroscopy, we have detected hydrogen and eleven metals in the atmosphere of the white dwarf. The origin of these metals is almost certainly the circumstellar disc of dusty and gaseous debris from a tidally-disrupted planetesimal, accreting at a rate of $1.6\times10^{10}\,\mathrm{g\,s^{-1}}$. Studying the chemical abundances of the accreted material demonstrates that the planetesimal had a composition similar to the Earth, dominated by rocky silicates and metallic iron, with a low water content. The mass of metals within the convection zone of the white dwarf corresponds to an asteroid of at least $\sim$130--170\,km in diameter, although the presence of ongoing accretion from the debris disc implies that the planetesimal was probably larger than this. While a previous abundance study of the accreted material has shown an anomalously high mass fraction of carbon (15\,percent) compared to the bulk Earth, our independent analysis results in a carbon abundance of just 2.5\,percent. Enhanced abundances of core material (Fe, Ni) suggest that the accreted object may have lost a portion of its mantle, possibly due to stellar wind stripping in the asymptotic giant branch. Time-series spectroscopy reveals variable emission from the orbiting  gaseous disc, demonstrating that the evolved planetary system at SDSS\,J0845+2257 is dynamically active. 
\end{abstract}


\begin{figure*}
    \centering
    \includegraphics[width=15cm]{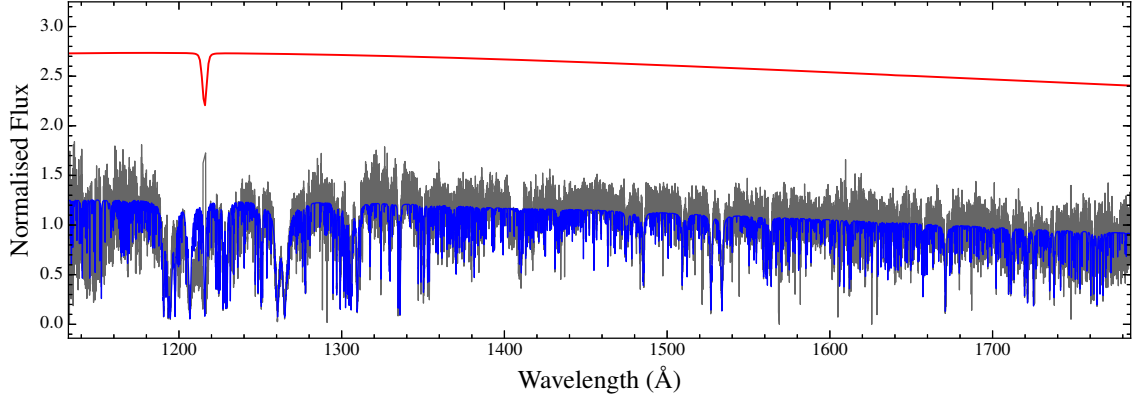}
    \caption{{\em HST}/COS FUV spectrum of SDSS\,J0845+2257 obtained on 2010 April 01 (grey), with the model fit used to determine the abundances of the accreted metals (blue). Plotted in red is a model spectrum for a white dwarf with the same atmospheric parameters, but no metals. The extremely large number of metal absorption lines in the spectrum of SDSS\,J0845+2257 is successfully reproduced by the model fit, with the exception of the two sections shown in Figure \ref{fig:bits}.\protect\label{fig:fuv}}
\end{figure*}

\begin{figure*}
    \centering
    \includegraphics[width=15cm]{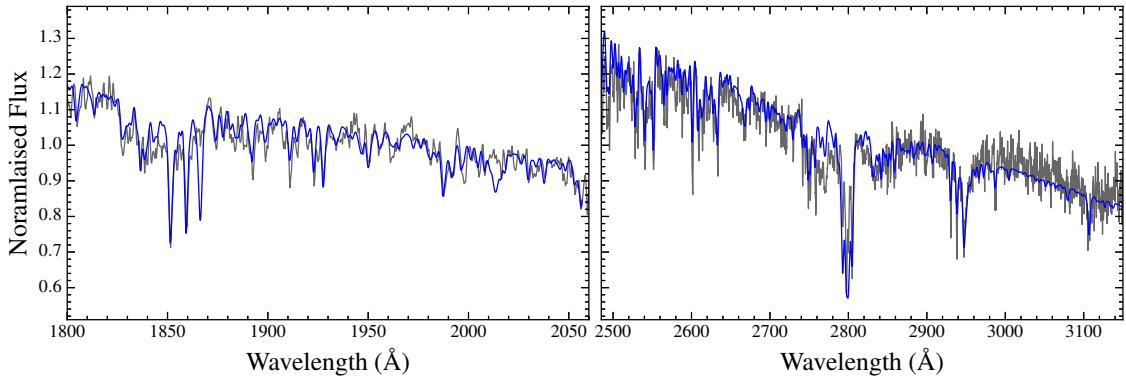}
    \caption{{\em HST}/COS NUV spectrum of SDSS\,J0845+2257 obtained on 2010 March 31 (grey), with the model fit used to determine the abundances of the accreted metals (blue). The model under-predicts the \Ion{Mg}{ii} 2790\AA\ triplet, possibly due to emission from the gaseous debris disc (\S\,\ref{sec:gas}).      \protect\label{fig:nuv}}
\end{figure*}

\begin{figure*}
    \centering
    \includegraphics[width=15cm]{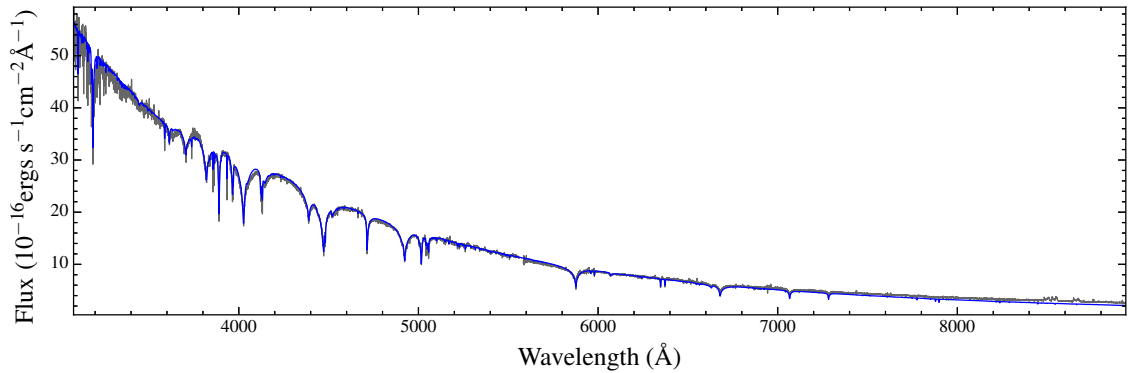}
    \caption{Full optical spectrum of SDSS\,J0845+2257 from the UVB and VIS arms of X-shooter (grey) together with the model fit (blue) used to calculate the atmospheric parameters and the metal abundances. A telluric correction was applied using the X-shooter spectral library \citep{chenetal14-1}. The model is over-plotted using just one scaling factor for the entire spectrum, demonstrating the excellent flux calibration of the X-shooter data.   \protect\label{fig:vis}}
\end{figure*}

\begin{figure}
    \centering
    \includegraphics[width=7.5cm]{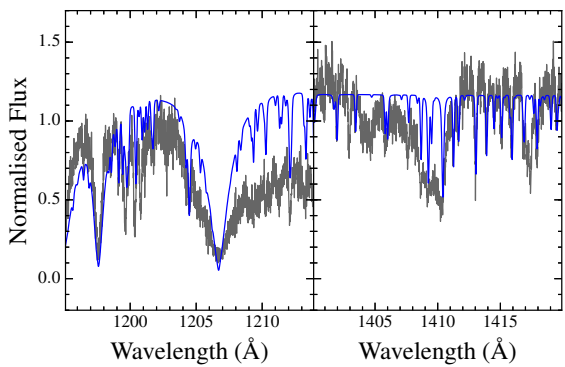}
    \caption{The model fit to the FUV spectrum under-predicts the red wing of the 1206\,\AA\ \Ion{Si}{iii} resonance line and what appear to be Si lines around 1400\,\AA. The reason for the poor fit in these areas is unknown, but a similar feature is seen in GALEX\,1931+0117 \citep[see fig. 4 in][]{gaensickeetal12-1}.    \protect\label{fig:bits}}
\end{figure}

\section{Introduction}
\label{sec:intro}
Since the discovery of calcium absorption lines in the atmosphere of van Maanen 2 \citep{vanmaanen17-1,weidemann60-1}, an explanation has been sought for the metal pollution observed in many white dwarfs. The high surface gravity ($\log g$) of white dwarfs implies that all metals will sink out of their atmospheres on time-scales that are short relative to the cooling age, leaving pristine hydrogen or helium atmospheres. However, metal pollution is now thought to be present at 25--50\,percent of all white dwarfs \citep{zuckermanetal03-1, zuckermanetal10-1, koesteretal14-1,barstowetal14-1}. Whilst radiative levitation can explain the observed metals in some white dwarfs with effective temperatures ($T_{\mathrm{eff}}$) above 20000\,K \citep{chayeretal95-1,chayer14-1,koesteretal14-1}, metal-polluted white dwarfs cooler than this must be currently accreting, or have recently accreted, material from an external source \citep{koester09-1}.

Historically the source of this accretion was thought to be the interstellar medium \citep{aannestadetal85-1,sionetal88-1}. The explanation that is now generally accepted developed from the infrared detection of a dusty debris disc, formed by the tidal disruption of an asteroid, around the white dwarf G29-38 \citep{zuckerman+becklin87-1,grahametal90-1,jura03-1}. Since then more than 30 similar discs have been discovered, and current estimates
suggest that 1--3\,percent of white dwarfs posses detectable dusty debris disks
\citep{farihietal09-1, girvenetal11-1, steeleetal11-1, barberetal14-1,rocchettoetal14-1}. Debris from these disrupted planetesimals accreting on the white dwarfs is now thought to be the dominant source of the atmospheric metals. In summary, these discoveries reveal the presence of evolved planetary systems at white dwarfs, offering an insight into the end-stages of planetary evolution. \cite{zuckerman14-1} noted that, in hindsight, the discovery of van Maanen 2 was therefore the first observational evidence of an extrasolar planetary system. 

In most white dwarfs pollution from only one or two elements, usually Ca and/or Mg, has been detected. However high--resolution spectroscopy of an increasing number of systems has revealed pollution by large number of metals, with a record of 16 species detected in the case of GD\,362 \citep{xuetal13-1}. Around a dozen white dwarfs show photospheric O, Mg, Si and Fe \citep{gaensickeetal12-1,dufouretal12-1,kleinetal11-1,juraetal12-1,farihietal13-2,xuetal14-1,raddietal15-1}. These four elements make up \textgreater 90\% of the bulk Earth \citep{allegreetal01-1}.

Consequently these systems provide a unique opportunity to study the bulk chemical composition of extrasolar planetary systems. Thus far, two main conclusions have been reached: 1. To zeroth order the chemical composition of the accreted debris is similar to that of the terrestrial planets in the Solar System \citep{zuckermanetal07-1,jura+young14-01}, with a distinct lack of volatile elements \citep{jura06-1,farihietal09-1}; 2. Within this overall similarity there is a large amount of diversity, with some systems showing evidence of differentiation \citep{gaensickeetal12-1}, post-nebula processing \citep{xuetal13-1} and water-rich asteroids \citep{farihietal13-2,raddietal15-1}.  

We present ultraviolet and times-series optical spectra of the metal-polluted DB white dwarf SDSS\,J084539.17+225728.0 (henceforth SDSS\,J0845+2257). This object was originally classified as a sdO subdwarf, Ton 345 \citep{greenetal86-1}. As part of a search for \Ion{Ca}{ii} 8600\,\AA\ emission lines among white dwarfs with SDSS spectra, \cite{gaensickeetal08-1} found that SDSS\,J0845+2257 was in fact a DB (helium atmosphere) white dwarf with a gaseous disc. Follow-up observations obtained with the William Hershel Telescope (WHT) revealed a significant change in the shape and strength of the \Ion{Ca}{ii} 8600\,\AA\ emission line profile, along with strong photospheric absorption lines from Ca, Si and Mg. 

\cite{juraetal14-1} presented a Keck/HIRES study detecting 11 metals in the atmosphere of SDSS\,J0845+2257. We extend these observations into the ultraviolet and carry out an  independent detailed study of the accreted material. We also present updated time-series observations of variable emission from the gaseous disc around SDSS\,J0845+2257 and an analysis of the {\em HST}/COS high-speed ultraviolet photometry. 

\begin{table*}
\centering
\caption{Log of observations of SDSS\,J0845+2257}
\begin{tabular}{lcccc}\\
\hline
Date & Telescope/ & Wavelength & Spectral & Total Exposure    \\
& Instrument & Range (\AA) & Resolution (\AA) & Time (s)\\ 
\hline
2004 December 19 & SDSS & 3794--9199 & 0.9--2.1 &3300\\
2008 January 02  & WHT/ISIS & 3577--8840 & 1.0--2.0&2400\\
2008 Apri1 03--5 & VLT/UVES & 3280--9460 & 0.02 & 17820\\
2008 January 08 & VLT/UVES & 3280--9460 & 0.02 & 11880\\
2009 April 09--11 & VLT/UVES & 3280--9460 & 0.02 & 17820\\
2010 March 31 & {\em HST}/COS & 2470--3150 & 0.01 & 7727\\
2010 April 01 & {\em HST}/COS & 1600--3150 & 0.39 & 14397\\
2010 April 02 & Gemini South/GMOS & 7540--9665 & 2.0 & 4980\\
2011 January 29 & VLT/X-shooter & 2990--24790 & 0.2--0.6 & 6990\\
2014 October 20 & VLT/X-shooter & 2990--24790 & 0.2--0.6 & 3731\\
\hline 
\end{tabular}
\label{tab:obs}
\end{table*}

\section{Observations}
\label{sec:obs}
Since the publication of \cite{gaensickeetal08-1} we have obtained deep spectroscopic observations of SDSS\,J0845+2257 in the ultraviolet with the {\em Hubble Space Telescope} ({\em HST}) and in the optical with the ESO Very Large Telescope (VLT) and Gemini South. Table \ref{tab:obs} provides a log of our observations.

{\em HST} observed SDSS\,J0845+2257 on 2010 March 31 and 2010 April 01 with the Cosmic Origins Spectrograph \citep[COS,][]{greenetal12-1}. Eight orbits were awarded under proposal  ID 11561 for a total exposure time of 2735\,s, 4994\,s and 14397\,s with the G130M, G160M and G230L gratings respectively. The spectra were processed with {\sc calcos 2.12}. Both the FUV (1120--1800\,\AA, Figure \ref{fig:fuv}) and NUV (1600--2060\,\AA, 2470--3150\,\AA, Figure \ref{fig:nuv}) spectra show a host of metal absorption lines, with only small amounts of continuum remaining.

Optical observations of SDSS\,J0845+2257 were obtained at the VLT on 2008 April 03--05, 2009 January 08 and 2009 April 09--11 with the Ultraviolet and Visual Echelle Spectrograph (UVES, \citealt{Dekkeretal00-1}) and again on 2011 January 29 and 2014 October 20 with X-shooter \citep{vernetetal11-1}. Total exposure times were 47520\,s with UVES and 10721\,s with X-shooter. 
These observations were reduced using the standard procedures within the {\sc reflex}\footnote{The {\sc reflex} software and documentation can be obtained from {\tt http://www.eso.org/sci/software/reflex/}} reduction tool developed by ESO. A heliocentric correction was applied to the {\sc reflex} outputs, and multiple exposures from the same night were combined.
The optical spectra show a large number of absorption lines, indicative of pollution from a variety of metals. The 8600\,\AA\ \Ion{Ca}{ii} emission line triplet is also clearly visible in all of the observations (see \S\,\ref{sec:gas}) .

On 2010 April 02 we obtained six spectra of SDSS\,J0845+2257 using the Gemini Multi-object Spectrograph on Gemini South \citep[GMOS,][]{hooketal04-1}. The observations were made in service mode, using the R831 grating and a 1\arcsec\ slit, which gave a wavelength range of 7540--9665\,\AA\ at a resolution of 2.0\,\AA. 
We requested the acquisition to be done in the $i$ band to ensure that the target was properly centred on the slit in the region of the \Ion{Ca}{ii} lines.
We reduced and extracted the spectra using the {\sc starlink} software packages {\sc kappa} and {\sc pamela}, and then applied the wavelength and flux calibration using {\sc molly}\footnote{{\sc molly} was written by T.\,R.~Marsh and is available from http://www.warwick.ac.uk/go/trmarsh/software/.}. The wavelength calibration was derived from a CuAr arc exposure taken in the morning following the observations and adjusted according to the known wavelengths of strong night sky emission lines. The flux calibration was calculated using an observation of the spectroscopic standard star LTT3218. Finally, we combined the six individual exposures to give a single, high signal-to-noise spectrum with a combined exposure time of 4980s.
The strength of the 8600\,\AA\ \Ion{Ca}{ii} triplet in this spectrum is comparable to the other optical observations. 

\begin{figure}
    \centering
    \includegraphics[width=7.5cm]{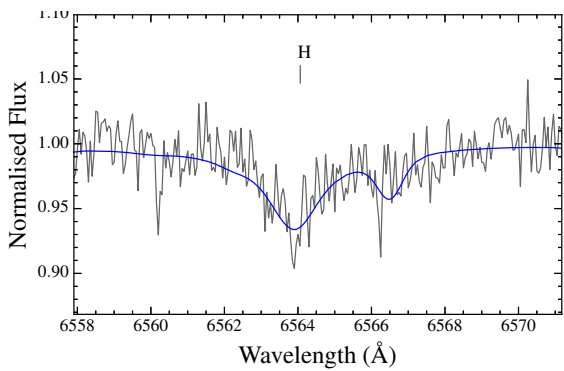}
    \caption{Section of our UVES spectrum showing a weak H$\alpha$ absorption line. Fitting an atmospheric model (blue) to the spectrum results in $\log$ (H/He)=$-5.10\pm0.50$.    \protect\label{fig:halpha}}
\end{figure}

\begin{figure}
    \centering
    \includegraphics[width=7.5cm]{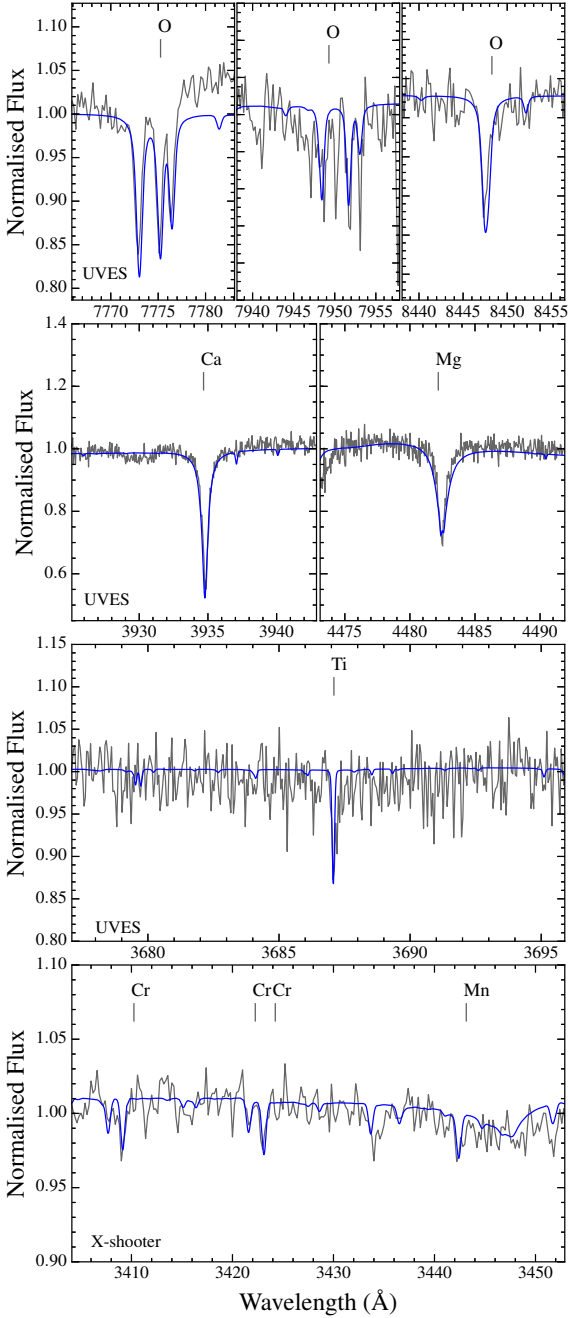}
    \caption{Normalised sections of X-Shooter and UVES spectra showing absorption lines from O, Ca, Mg, Ti, Cr and Mn, with the model atmosphere fit used to calculate the abundances over-plotted in blue. The section shown in the top-middle panel is affected by telluric lines, which are not reproduced by our model.  \protect\label{fig:optlines}}
\end{figure}

\begin{figure}
    \centering
    \includegraphics[width=7.5cm]{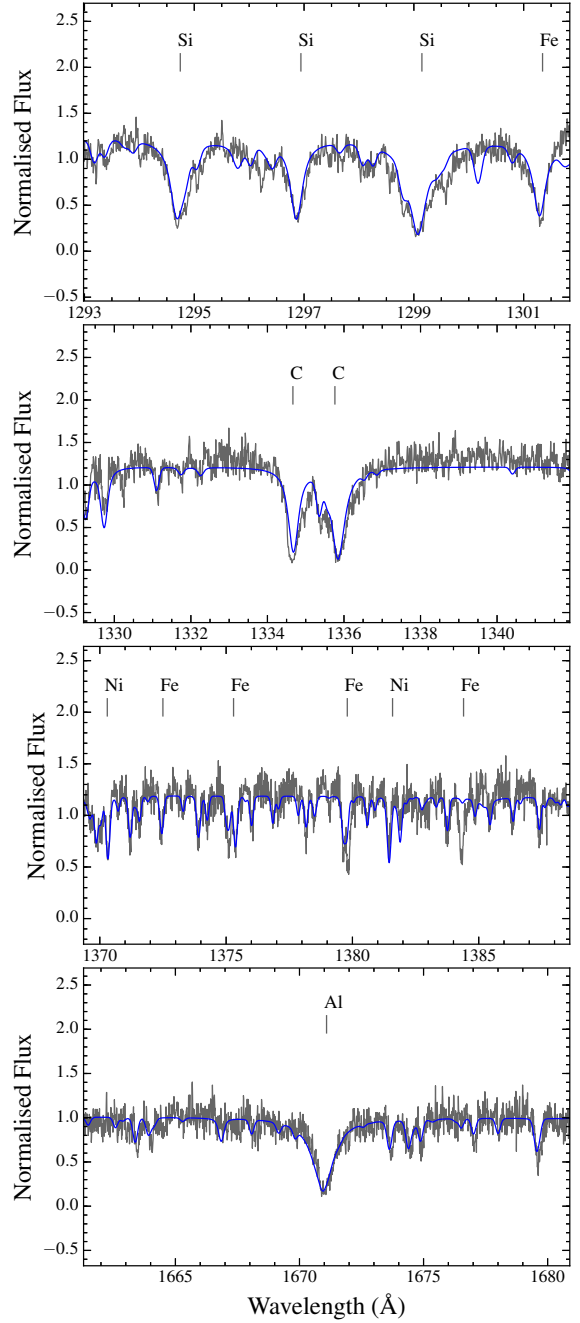}
    \caption{Sections of the COS FUV spectrum showing absorption lines from Si, C, Ni, Fe and Al, with the model atmosphere fit used to calculate the abundances over-plotted in blue. \protect\label{fig:fuvlines}}
\end{figure}

\section{White Dwarf Parameters}
\label{sec:params}

The spectroscopic determination of atmospheric parameters for hot DB
white dwarfs is extremely difficult, because the spectra hardly change
between $T_{\mathrm{eff}}=$20000--30000\,K. Using a fit to the SDSS photometry with a fixed $\log g=8.00$, we obtain $T_{\mathrm{eff}}=19850\pm600$\,K for a pure He grid, $T_{\mathrm{eff}}=19890\pm620$\,K for a grid with $\log$ (H/He)$=
{-}5.0$, and $T_{\mathrm{eff}}=19880\pm600$\,K for a grid with our final metal
abundances. The increase of the free electron density, which is the
major effect from the inclusion of metals in cooler DB models, has no
significant effect at these temperatures. There is, however, a
noticeable blanketing effect from the strong metal lines in the
ultraviolet. This increases the overall flux in the optical range, but with the
same factor in all SDSS band passes. It therefore has no effect on
the atmospheric parameter determination.

A possible difficulty when using the photometry for faint and distant
objects is the interstellar reddening. The maximum reddening in the
direction of SDSS\,J0845+2257 is E(B-V) = 0.0268 \citep{schlafly+finkbeiner11-01}. If we apply this value and repeat the fitting
with the dereddened photometry, the result is a much higher temperature
of $T_{\mathrm{eff}}=24320\pm920$\,K. From the solid angle obtained through this fit, and
assuming a radius of the white dwarf which corresponds to $\log g=8.0$, we can obtain the distance to the white dwarf and estimate the true reddening. We find a distance of 117 pc and a vertical distance above the galactic plane of 67 pc. Using the algorithm of \citet[see also \citealt{genest-beaulieu+bergeron14-1}]{tremblayetal11-2}, 
we estimate a negligible reddening of $E(B-V) = 0.001$, and thus the lower $T_{\mathrm{eff}}$ is secure.

The atmospheric parameters from the photometry are confirmed by a fit to the SDSS spectrum,
which results in $T_{\mathrm{eff}}=19800\pm70$\,K, $\log g = 8.16\pm0.02$ for the pure He
grid, and $T_{\mathrm{eff}}=19780\pm70$\,K, $\log g= 8.18\pm0.02$ for the grid with $\log$ (H/He) $=-5.0$. A final test is offered by the absolute calibration of the {\em HST}
spectra. Using the solid angle from the photometry, the effective
temperature is confined to $T_{\mathrm{eff}}=19750\pm250$\,K by the comparison of the
predicted ultraviolet fluxes with the observations. Our final compromise for
the atmospheric parameters is the spectroscopic fit with the $\log$ (H/He) $=-5$
grid, with enlarged errors:
\begin{displaymath}
       T_{\mathrm{eff}}=19780\pm250\,\mathrm{K} \ \ \ \    \log g = 8.18\pm0.20
\end{displaymath}

The hydrogen abundance is consistent with a fit to the H$\alpha$ line found in the UVES spectrum (Figure \ref{fig:halpha}). 

\cite{juraetal14-1} rely entirely on a fit to the SDSS photometry, obtaining
$T_{\mathrm{eff}}=19535\pm700$\,K for a pure He grid and $T_{\mathrm{eff}}=18700\pm700$\,K for a grid with the
observed metal abundances. The surface gravity cannot be obtained from
the photometry so they fixed $\log g=8.00$. We have computed a model adopting the parameters of \cite{juraetal14-1} and multiplied it with the solid angle obtained from their
photometric fit. Their lower $T_{\mathrm{eff}}$ under-predicts the COS ultraviolet fluxes by 15--23\,percent. Whilst a temperature
difference of 1000\,K would not normally have a large effect on the metal abundances, in this case all elements heavier than
oxygen are in a transition from first to second ionization stage so the predicted lines in our, hotter, model are
significantly weaker. We thus obtain larger abundances for all
elements, with the exception of carbon. Our carbon abundance is based on the COS ultraviolet spectroscopy, which contains more and stronger carbon lines than the optical data analysed by \cite{juraetal14-1}. An independent model atmosphere analysis of the COS ultraviolet spectra of SDSS\,J0845+2257 carried out by P. Dufour (private communication), adopting our atmospheric parameters, confirms the low photospheric carbon abundance.

Using our results for $T_{\mathrm{eff}}$ and $\log g$, and the helium atmosphere models of Bergeron and collaborators\footnote{{\tt http://www.astro.umontreal.ca/${\sim}$bergeron/CoolingModels}, based on \cite{holberg+bergeron06-1}, \cite{kowalski+saumon06-1}, \cite{tremblayetal11-2} and \cite{bergeronetal11-1}.}, we find a cooling age $T_{\mathrm{cool}}\approx100$\,Myr, white dwarf mass $\Mwd=0.679\,\Msun$ and radius $\Rwd=0.011\,\Rsun$. Starting from a Rossland optical depth $\tau_{\mathrm{r}}=100$ and integrating the envelope equations downwards using the equation
of state of \cite{saumonetal95-1}, we find a convection zone mass of $\log (M_{\mathrm{cvz}}/\Mwd)=-8.4$. Using the initial-to-final mass relationships of \cite{gesickietal14-1}, \cite{casewelletal09-1} and \cite{catalanetal08-2} we find a progenitor mass of $\sim2.7\,\Msun$, similar to most metal-polluted white dwarfs \citep{jura+xu12-1,koesteretal14-1} and equivalent to an A-type star progenitor.

\begin{table}
\centering
\caption{Measured atmospheric abundances, computed diffusion time-scales and inferred metal accretion rates in the atmosphere of SDSS\,J0845+2257.}
\begin{tabular}{llcr}
\hline
Element & log $n(\mathrm{Z})/n(\mathrm{He})$ & $\tau_\mathrm{diff} (10^4 \mathrm{yr})$ & $\dot M~[\mathrm{g\,s^{-1}}]$ \\
\hline
1 H   & $-5.10\pm0.50$ &  n/a   & n/a \\
6 C   & $-4.90\pm0.20$ & $1.5$  & $4.1\times10^8$\\
7 N   & ${\leq}-6.30$  & $1.5$  & ${\leq}1.9\times10^7$\\
8 O   & $-4.25\pm0.20$ & $0.97$ & $3.8\times10^9$\\
12 Mg & $-4.70\pm0.15$ & $1.2$  & $1.6\times10^9$\\
13 Al & $-5.70\pm0.15$ & $1.2$  & $1.9\times10^8$\\
14 Si & $-4.80\pm0.30$ & $1.2$  & $1.5\times10^9$\\
16 S  & ${\leq}-5.40$  & $1.2$  & ${\leq}4.3\times10^8$\\
20 Ca & $-5.95\pm0.10$ & $0.97$ & $1.9\times10^8$\\
21 Sc & ${\leq}-7.70$  & $0.65$ & ${\leq}5.7\times10^6$\\
22 Ti & ${\leq}-7.15$  & $0.59$ & ${\leq}2.4\times10^7$\\
24 Cr & $-6.40\pm0.30$ & $0.53$ & $1.6\times10^8$\\
25 Mn & $-7.00\pm0.40$ & $0.49$ & $4.6\times10^7$\\
26 Fe & $-4.60\pm0.20$ & $0.87$ & $6.6\times10^9$\\
28 Ni & $-5.65\pm0.30$ & $0.44$ & $1.2\times10^9$\\
Total &                &        & $1.6\times10^{10}$\\
\hline 
\end{tabular}
\label{tab:abs}
\end{table}

\begin{table}
\centering
\caption{Mass fractions of the accreted debris in the convection zone of SDSS\,J0845+2257 and in the bulk Earth \citep{allegreetal01-1}. In the early-phase/instantaneous approximation the mass fractions are calculated using the atmospheric abundances, whilst in steady-state the inferred accretion rates are used. 
The differences between the two approximations are small for most elements.}
\begin{tabular}{lccc}
\hline
Element & \multicolumn{3}{c}{Percentage by mass} \\
 & \multicolumn{2}{c}{SDSS\,J0845+2257} & Bulk Earth  \\
 & Early-Phase & Steady-State & \\
\hline
6 C   & $4.0\pm1.8$   & $2.5\pm1.2$   & 0.17 \\
8 O   & $23.8\pm11.0$ & $23.4\pm10.8$ & 32.4 \\
12 Mg & $12.7\pm4.4$  & $9.9\pm3.4$   & 15.8 \\
13 Al & $1.4\pm0.5$   & $1.2\pm0.4$   & 1.5  \\
14 Si & $11.8\pm8.1$  & $9.3\pm6.4$   & 17.1 \\
20 Ca & $1.2\pm0.3$   & $1.2\pm0.3$   & 1.62 \\
24 Cr & $0.55\pm0.38$ & $0.99\pm0.68$ & 0.42 \\ 
25 Mn & $0.15\pm0.13$ & $0.28\pm0.26$ & 0.14 \\
26 Fe & $37.3\pm17.1$ & $40.8\pm18.8$ & 28.8 \\
28 Ni & $3.4\pm2.4$   & $7.5\pm5.2$   & 1.69 \\
Other & $3.7$         & $2.9$         & 1.08 \\            
\hline 
\end{tabular}
\label{tab:pers}
\end{table}

\section{Accretion of Planetary Material}
\label{sec:accretion}
The ultraviolet and optical spectra (Figures \ref{fig:fuv}, \ref{fig:nuv} and \ref{fig:vis}) of SDSS\,J0845+2257 show metal absorption lines from a variety of elements, from which we can confidently measure the atmospheric abundances for hydrogen and 10 metals and place an upper limit on 4 further metals.

The spectra show many dozens to hundreds of lines of Mg, Si, Ca and Fe, so the measured abundances for those elements are fairly secure. The Ni abundance is confirmed by the clear and moderately strong lines at 1317\,\AA\ and 1370\,\AA. Our carbon abundance is based on more than 15
strong \Ion{C}{i} and \Ion{C}{ii} lines between 1270\,\AA\ and 1470\,\AA\ and thus also fairly robust. The O abundance relies on the 7777\,\AA, 7949\,\AA\ and 8448\,\AA\ lines shown in Figure \ref{fig:optlines}. As with \cite{dufouretal12-1}, we find the abundance obtained using the 7949\,\AA\ absorption line to be highly discrepant with the other two lines. As the available atomic data for this line is limited, we have chosen to neglect this point, with the final abundance and error calculated from the remaining two lines. The 1152\,\AA\ and 1302\,\AA\ lines detected in the ultraviolet spectra allow a less precise measurement which is compatible with this result. 
     
One of the four metals for which we only present upper limits, Ti, is actually detected in our spectra. However there is disagreement between the abundance measurements from lines at 3686\,\AA\ and 
3762\,\AA, which fit the quoted abundance value, and the 3760\,\AA\ line, which is weaker than predicted by the model. Another line predicted by the model atmosphere, 3901\,\AA, is not observed. For these reasons we only feel confident to present an upper limit for the abundance of Ti.     

To study the composition of the progenitor object we must compute the relative abundance ratios of the elements being accreted into the white dwarf atmosphere. These are not identical to the ratios of the photospheric metal abundances derived above, as individual metals sink out of the He envelope on different diffusion time-scales. The diffusion time-scales are a function of the depth of the convection zone and the diffusion velocity of each element, both of which vary with $T_{\mathrm{eff}}$ \citep{koester09-1}. 

SDSS\,J0845+2257 has a shallow  (log $(M_{\mathrm{cvz}}/\Mwd)=-8.4$) convection zone and we assume that the accreted metals are homogeneously mixed, such that the relative ratios of elements near the bottom of the convection zone are the same in the photosphere.  We calculate individual diffusion time-scales for each element  (column 3 of Table \ref{tab:abs}) using the techniques described in \cite{koester09-1}\footnote{See updated values at {\tt http://www1.astrophysik.uni-kiel.de/\\~koester/astrophysics/astrophysics.html}.}, taking the  total mass of helium in the convection zone to be $M(\mathrm{He})_{\mathrm{cvz}}=2.605\times10^{-9}$\Msun. The accretion rates of each element onto the white dwarf are then proportional to the ratio of the the photospheric metal abundance to the diffusion time-scale, leading to the results in column 4 of Table \ref{tab:abs}.

The accretion/diffusion computations of \cite{koester09-1} assume that accretion has been ongoing for $\gtrsim$5 diffusion time-scales, reaching a steady-state between material diffusing out of the convection zone and accreting onto the white dwarf. This assumption is likely valid as a dusty debris disc is detected \citep{brinkworthetal12-1}, which is almost certainly the source of the metals. \cite{girvenetal12-1} showed that such discs have an estimated lifetime of several $10^5$\,yr. As the diffusion time-scales are only of order $10^4$\,yr it is reasonable to assume that a steady-state has been reached.  

However, we cannot exclude the possibilities that the debris disc formed recently and that accretion and diffusion have not reached a steady-state (the early-phase), or that the accretion rate may not be constant over sufficiently long time-scales. The presence of a gaseous component to the debris disc (\S\,\ref{sec:gas}) has been suggested to be the result of dynamical activity in the disc \citep{verasetal14-1}, which may affect the accretion rate. As the lifetime of the gaseous component is likely to be short relative to the diffusion time-scale \citep{wilsonetal14-1}, its presence may be a sign of recent changes in the accretion rate. In this case, known as the  instantaneous assumption, the relative chemical abundances in the accreted material will match the photospheric abundances. \cite{juraetal14-1} based their analysis of the debris composition on the instantaneous assumption, as at the $T_{\mathrm{eff}}=18700$\,K used in their analysis the diffusion time-scales are much longer, making it much less likely for the accretion to have reached a steady-state.    

Table \ref{tab:pers} shows the relative mass fractions of the elements under both the steady-state and early-phase/instantaneous scenarios. We find that the choice of accretion scenario has only a small effect on the elemental mass fractions, with the possible exceptions of C and Ni. Our discussion below of the debris composition is based on the steady-state assumption, but we note where the differences between steady-state and early-phase accretion are significant. 

\subsection{Total Accretion Rate and Mass of the Parent Body}
\label{sec:tot}
At $1.6\times10^{10}\,\mathrm{g\,s^{-1}}$, SDSS\,J0845+2257 has one of the highest inferred accretion rates detected at a metal-polluted white dwarf, with the observed elements representing the bulk of those making up the Earth (Table \ref{tab:pers}). Any additional undetected elements make up only trace amounts, so we can therefore draw reliable conclusions about the bulk abundances of the accreted material. 

The total mass of metals calculated to be in the convection zone is $4.9\times10^{21}$\,g, setting a lower limit on the mass of the accreted object (or objects). Assuming a rock-like density of $\rho\approx2{-}4\,\mathrm{gcm^{-3}}$, the mass is equivalent to a spherical object with a ${\sim}$130--170\,km diameter. However, detection of circumstellar gas \citep{gaensickeetal08-1} and dust \citep{brinkworthetal12-1} implies that accretion is still ongoing. If a large proportion of the debris is still in the disc, and/or has already sunk out of the convection zone of the white dwarf, then the total mass of the progenitor object may have been significantly higher than that currently present in the convection zone. 

\begin{figure}
    \centering
    \includegraphics[width=8.5cm]{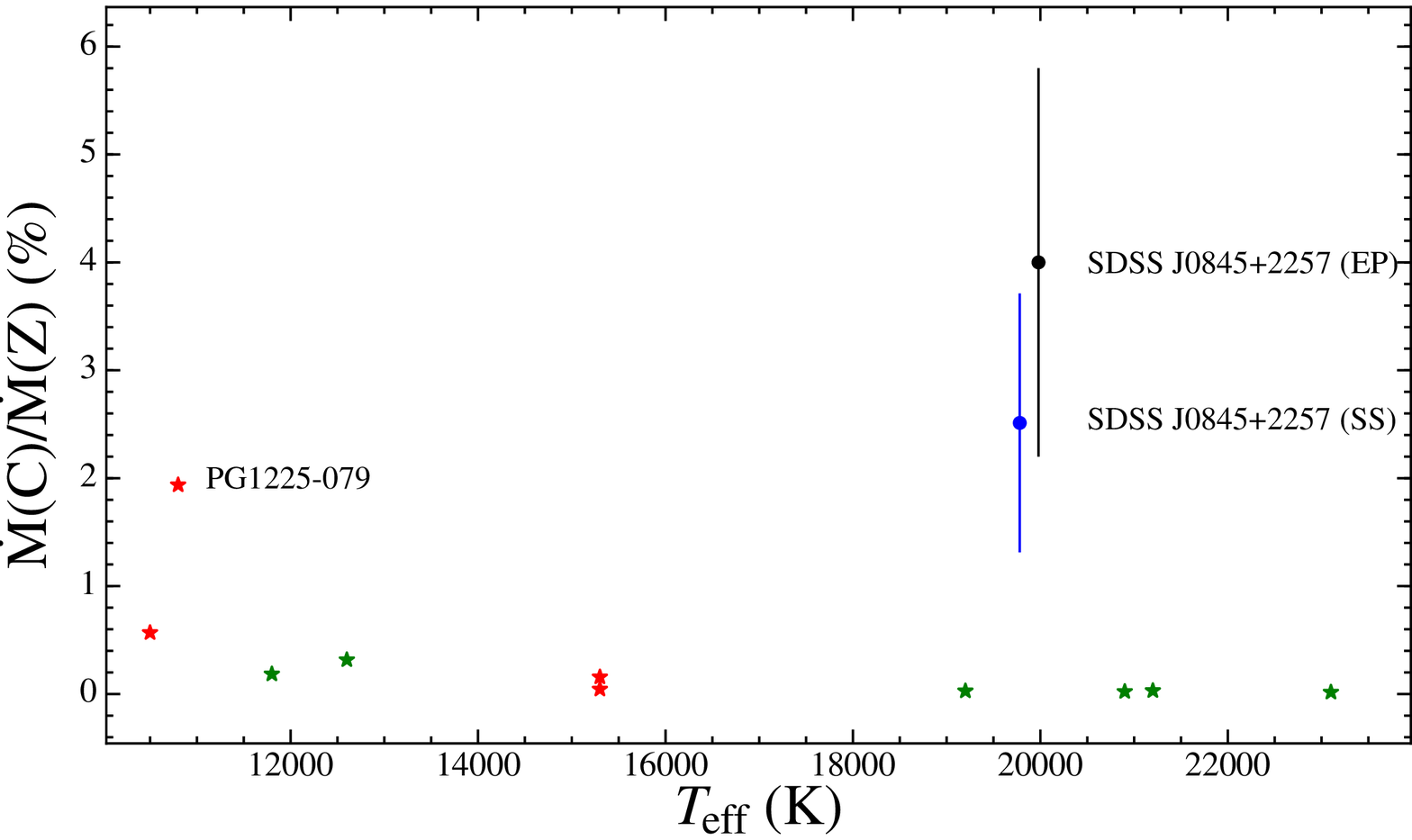}
    \caption{Mass fraction of carbon in the accreted material analysed in the most heavily polluted white dwarfs \citep{zuckermanetal11-1,kleinetal11-1,dufouretal12-1,gaensickeetal12-1,farihietal13-2,xuetal13-1,xuetal14-1}. Both the early-phase (black) and steady-state (blue) results for SDSS\,J0845+2257 are shown, with the marker for the early-phase offset by +200\,K for clarity. Whilst the carbon fraction is low and roughly constant for both DA (green star symbols) and DB (red star symbols) white dwarfs regardless of $T_{\mathrm{eff}}$, there are two clear outliers: PG\,1225-079 \citep{xuetal13-1} and SDSS\,J0845+2257.          
       \protect\label{fig:carbon}}
\end{figure}

\subsection{Carbon}
\label{sec:carbon}
Figure \ref{fig:carbon} shows that the mass fraction of carbon in the accreted material for nearly all of the most heavily polluted white dwarfs is ${\lesssim}$\,0.5\,percent, regardless of temperature, for both hydrogen (DA) and helium (DB) atmospheres. However two DB white dwarfs, PG\,1225-079 \citep{xuetal13-1} and SDSS\,J0845+2257, have much higher carbon abundances. In PG\,1225-079, which has a carbon mass fraction of 1.9\,percent, \cite{xuetal13-1} found no exact match of any solar system object and suggested that the accretion may have been caused by two or more objects with different compositions. \cite{juraetal14-1} found a carbon mass fraction of 15\,percent in SDSS\,J0845+2257, which, assuming that all of the carbon was accreted, is consistent with that of Interplanetary Dust Particles (IDPs). As IDPs are too small to make up the high inferred accretion rate, \cite{juraetal14-1} suggested that the disrupted rocky body was a Kuiper-belt analogue which lost its water content during the post main-sequence evolution of the system. Although our diffusion calculations still return an unusually high carbon abundance (2.5\,percent in steady-state, 4.0\,percent in early-phase), it is significantly lower than that found by \cite{juraetal14-1} and thus does not match the abundances found in IDPs. The large discrepancy between the carbon mass fractions derived from the two studies arises from the combination of a lower photospheric abundance of carbon determined from our COS ultraviolet spectroscopy (\S\,\ref{sec:params}) and a systematically higher abundance of heavier elements. However, although the detailed results of the two studies differ, the fundamental conclusion of an unusually large abundance of carbon in the of photosphere of SDSS\,J0845+2257, when compared to other debris-polluted white dwarfs, remains. We consider several alternative possible explanations for the excess carbon.

Firstly, \cite{verasetal14-2} speculated that a small fraction of the debris at white dwarfs could be made up of exo-Oort cloud comets, which do have a substantial carbon fraction  \citep{jessbergeretal88-1}. However, comets have a much higher water content than that observed here (\S\,\ref{sec:oxygen}) and cannot supply enough mass to explain the high rate and total amount of accreted metals. 

Another scenario that could explain the high carbon abundance in SDSS\,J0845+2257 is based on results from terrestrial seismology, which have shown that a portion of the Earth's core must be made up of less dense material than the majority Fe and Ni \citep{allegreetal95-1}. Carbon has been suggested as a possible candidate for this material \citep{poirier94-1,zhang+yin12-1}. The enhanced levels of known core elements (Table \ref{tab:pers}, \S\,\ref{sec:core}) at SDSS\,J0845+2257 could therefore explain the increased carbon abundance. However, the similarly enhanced level of core material detected in PG\,0843+516 \citep{gaensickeetal12-1} was not accompanied by an increase in the carbon mass fraction.

An alternative explanation is that the carbon present in the atmosphere of SDSS\,J0845+2257 has not been accreted from planetary debris, but is instead primordial to the white dwarf. A number of DB white dwarfs have been observed with atmospheric pollution from carbon, but no other metals \citep{provencaletal00-1,koesteretal14-2}. These white dwarfs span a range of $T_{\mathrm{eff}}$ and no single explanation (in the absence of accretion) can currently explain the high carbon abundances. At temperatures $\lesssim16000$\,K convection can dredge up carbon from the core into the helium layer \citep{koesteretal82-2}, although some DBs have been observed with carbon abundances much lower than predicted by this model \citep{desharnaisetal08-1}. We note that the maximum contamination by dredge-up is thought to occur near 12000\,K \citep{pelletieretal86-1}, providing an entirely plausible source for the carbon detected in PG\,1225-079 \citep[$T_{\mathrm{eff}}=10800$\,K,][]{xuetal13-1}. The carbon in hotter (\textgreater20000\,K) white dwarfs, which cannot be explained by the classical dredge up model, has been postulated to be raised by a weak stellar wind \citep{fontaine+brassard05-1}, but there is currently no working model for the necessary wind acceleration. Both the dredge-up and stellar wind models predict a very low carbon abundance in the temperature range ${\sim}$17000--20000\,K.  

However, the efficiency of photospheric carbon pollution by dredge-up
depends not only on the depth of the convection zone, but also on the
total mass of the helium layer \citep{weidemann+koester95-1}. The bulk of the DQ white dwarfs can be modelled
with helium envelopes of $\simeq10^{-3}-10^{-2}\,M_\odot$ \citep[e.g.][]{fontaine+brassard05-1}, which is in agreement
with the helium masses predicted by stellar evolution models \citep[e.g.][]{lawlor+macdonald06-1}. However, a number of DB
white dwarfs \citep{koesteretal14-2}, cooler DQ \citep{dufouretal05-1,koester+knist06-1}, as well
as the recently discovered hot DQ white dwarfs \citep{dufouretal07-1,dufouretal08-1} have higher carbon
abundances than predicted by the canonical dredge-up scenario, which can
be explained with thinner helium layers \citep{althausetal09-1,koesteretal14-2}. Assuming that the unusually high carbon abundances in SDSS\,J0845+2257 are a result of dredge-up would require
a thin ($\log$ (M$_\mathrm{He}$/\Mwd)$\sim$\,-4.6) helium layer. In the
light of the ongoing discussion of carbon abundances in helium
atmosphere white dwarfs, and in the absence of a robust sample of
helium layer measurements for white dwarfs, we conclude that dredge-up
is a plausible origin of the photospheric carbon in SDSS\,J0845+2257.

An independent evaluation of the origin of the atmospheric carbon may be provided by the models of \cite{hartmannetal14-1}, which predict that even a relativity small mass fraction of carbon ({\textgreater}10$^{-4}$) in the circumstellar debris should lead to emission from gaseous \Ion{C}{ii} at 8683\,\AA\ and 8697\,\AA. Given that we do detect emission from gaseous \Ion{Ca}{ii} (\S\,\ref{sec:gas}), the non-detection of \Ion{C}{ii} could be further evidence that the carbon is primordial to the white dwarf. \cite{hartmannetal11-1} applied the same model to the gaseous disc around SDSS\,J1228+1040, again finding that the absence of \Ion{C}{ii} 8683\,\AA\ and 8697\,\AA\ emission features in spectra obtained by \cite{gaensickeetal06-3} requires a low ($\lesssim$0.5\,percent) carbon mass fraction in the gas. \cite{gaensickeetal12-1} showed that SDSS\,J1228+1040 has an extremely low photospheric carbon abundance, unambiguously demonstrating that the debris disc in this system is indeed strongly carbon-depleted.

\begin{figure}
    \centering
    \includegraphics[width=8.cm]{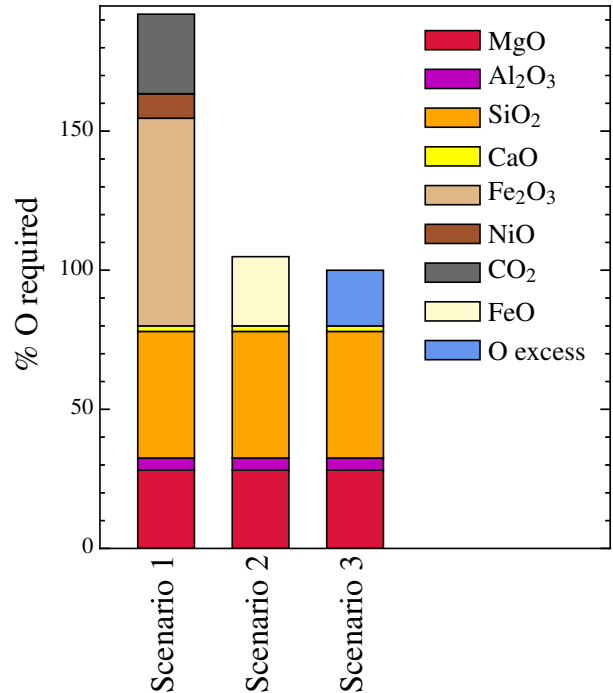}
    \caption{Oxygen budgets of the debris at SDSS\,J0845+2257, with three scenarios for different contributions of oxygen carriers: 1. All Fe as Fe$_2$O$_3$, CO$_2$ and NiO present; 2. 50\,\% of Fe as FeO; 3. All Fe is metallic, making no contribution to the O budget. The third model is the only one to produce a marginal oxygen excess, suggesting that very little water was present in the parent body.
       \protect\label{fig:obgt}}
\end{figure}
     
\begin{figure*}
    \centering
    \includegraphics[width=15cm]{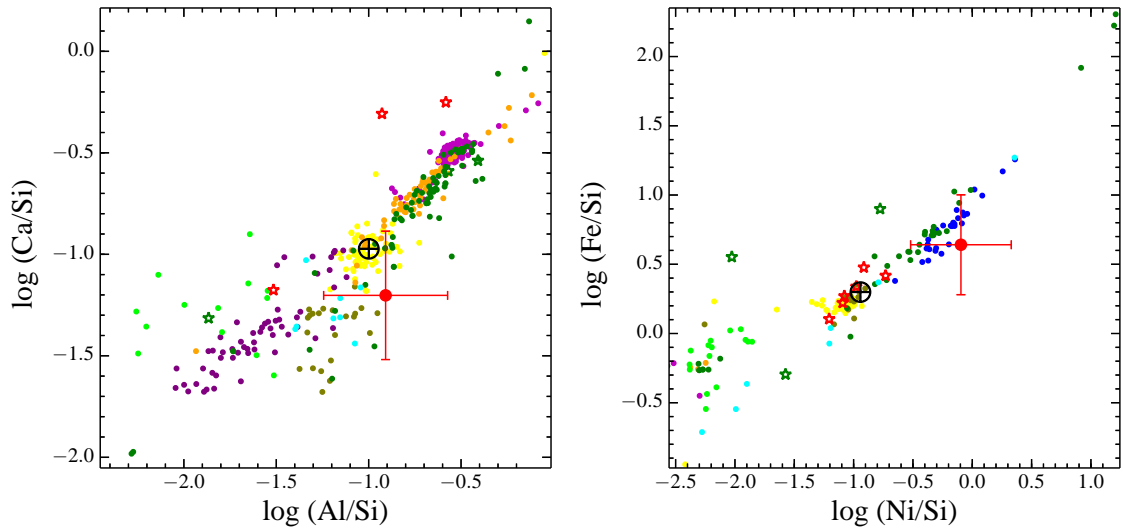}
    \caption{Comparison of the abundance of Al and Ca (left) and Fe and Ni (right) in SDSS\,J0845+2257 (red marker) with the bulk Earth ($\bigoplus$) \citep{mcdonough00-1} and the allasite (blue), mesosiderite (dark green), IAB (cyan), urelite (light green), ordinary chrondrite  and carbonaceous chrondrite (yellow), enstatite chrondrite (green-brown), Howardite (orange), Eucrite (magenta) and Diogenite (purple) meteorites \citep{nittleretal04-1}. The most heavily metal-polluted white dwarfs are plotted as green (DAZ) and red (DBZ) stars \citep{zuckermanetal11-1,kleinetal11-1,dufouretal12-1,gaensickeetal12-1,xuetal14-1,xuetal13-1,farihietal13-2,xuetal14-1}.  All abundances are normalised to Si. \protect\label{fig:ratios}}
\end{figure*}

\subsection{Oxygen}
\label{sec:oxygen}
Measurements of the oxygen abundance in the accreting white dwarf GD\,61 by \cite{farihietal13-2} suggested that a large fraction of the debris was made of water. To estimate the water content in the debris around SDSS\,J0845+2257 we follow the procedure of \cite{farihietal13-2}, assessing first oxygen carriers other than water. We assume that all of the accreting Mg, Al, Si, and Ca is bound into MgO, Al$_2$O$_3$, SiO$_2$ and CaO. If the debris is a fragment of a differentiated object then the Fe and Ni content may be split into oxides from the mantle and metallic Fe and Ni from the core, so we present three scenarios for the remaining elements: 1. A conservative scenario where all of the Fe is bound into Fe$_2$O$_3$, Ni into NiO and C into CO$_2$; 2. An intermediate scenario where C is primordial (see \S\,\ref{sec:carbon}), the contribution from Ni is negligible and half the Fe is present as FeO; 3. A mantle depleted scenario, discussed further in \S\,\ref{sec:core}, where there is no contribution from Fe, Ni or C.       
Only the third of these scenarios produces an oxygen excess, indicating that the accreted material was likely very dry regardless of the relative contributions from mantle and core material, as well as from carbon. This result is consistent with the low H abundance detected in the atmosphere (Table \ref{tab:abs}), and is not significantly altered under the assumption of early-phase or instantaneous accretion. Scenario 1 shows that there is insufficient oxygen to account for all of the potential carriers, an indicator that a large fraction of the Fe in the progenitor object was indeed metallic. 

\subsection{Refractory Lithophiles}
\label{sec:liths}
Figure \ref{fig:ratios} compares the abundance of Ca and Al with respect to Si. These elements are two of the main refractory lithophiles: elements which sublimate only at very high temperatures and are therefore found mainly in the mantle and crust of differentiated objects \citep{grossman72-1}. The ratio of these two elements is nearly constant in most solar system bodies, such that there is a linear correlation between Ca/Si and Al/Si. The Ca and Al accreting onto SDSS\,J0845+2257, as well as onto the other heavily polluted white dwarfs, also falls onto or near this line. This indicates that similar geochemical processes are taking place in these systems, and strengthens the case that analysing the accreted debris in the white dwarf photosphere provides a reliable representation of the chemical composition of a terrestrial object at SDSS\,J0845+2257. 

\begin{figure}
    \centering
    \includegraphics[width=8.5cm]{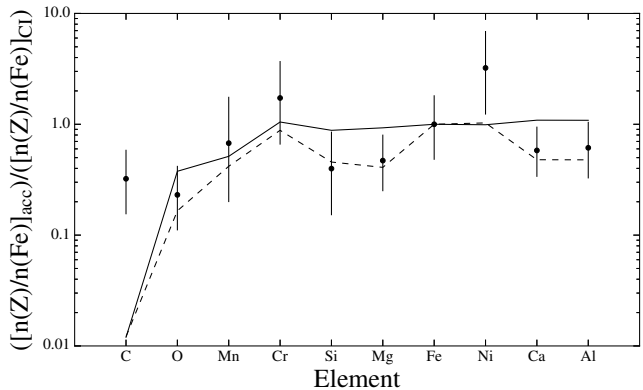}
    \caption{Elemental abundances of the accretion onto SDSS\,J0845+2257 relative to the abundance in CI chondrites, both normalised to Fe (black dots). The condensation temperature of the elements increases to the right. The solid line shows the abundances of the bulk Earth, whilst the dotted  line shows a `wind stripped' model \citep{melisetal11-1} in which approximately 15\,percent of the silicate Earth has been removed.         
       \protect\label{fig:strip}}
\end{figure}

\subsection{Iron and Nickel}
\label{sec:core}
Table \ref{tab:pers} shows that the dominant element in the debris at SDSS\,J0845+2257 is Fe, with a mass fraction of 40.8\,percent, substantially higher than in the Earth. Ni, which is the other major component in the Earth's core \citep{mcdonough00-1}, is also enhanced relative to the Earth. Figure \ref{fig:ratios} shows the linear relationship between Fe/Si and Ni/Si found in the solar system bodies. Whilst the heavily polluted DB white dwarfs also fall on this line, the DA white dwarfs are less confined. 
SDSS\,J0845+2257 is close to the trend but has a higher Ni/Fe ratio ($\approx0.2$) than the Earth ($0.06$). In the early-phase model the Ni/Fe ratio drops to $\approx0.09$.

The high Fe and Ni abundances suggest that the progenitor of the debris at SDSS\,J0845+2257 may have been a fragment of a larger, differentiated planetesimal with a relatively large core. This could be evidence that some of the processes thought to be responsible for Mercury's large core, such as partial volatilization \citep{cameron85-1} or iron/silicate fractionation \citep{weidenschilling78-1}, also occur in extrasolar planetary systems. However, it is likely that any planet close enough to its star for these processes to occur would have been engulfed during the red giant phase, unless it migrated outwards after formation.

\cite{melisetal11-1} proposed an alternative model in which a planetesimal is eroded by the intense stellar wind during the asymptotic giant branch (AGB). Figure \ref{fig:strip} shows this model applied to SDSS\,J0845+2257. Following the technique detailed in \cite{melisetal11-1}, we show the abundances of the material accreted onto the white dwarf (Table \ref{tab:abs}) relative to the abundances in the CI chondrites \citep{lodders03-1}, both of which are normalised to Fe. Combining the abundances for the core and silicate Earth from \cite{mcdonough00-1} we normalise the abundances in the bulk Earth in the same way (solid black line). We then remove $\sim$15\% of the silicate material (dotted line in Figure \ref{fig:strip}), simulating the erosion of the mantle of a terrestrial object by stellar wind. We neglect any contribution from the crust, as it makes up only ${\approx}$0.5\,percent of the silicate Earth \citep{allegreetal95-1}. As can be seen in Figure \ref{fig:strip} the dominant elements of both Earth and SDSS\,J0845+2257 (e.g. Si, O, Mg, Ca, Al) appear to support this model, whilst the trace elements (Mn, Cr) are consistent with both scenarios. The high abundances of C (if accreted, see \S\,\ref{sec:carbon}) and Ni remain unexplained, although Ni does become consistent with the model under the early-phase assumption. We note that, whilst \cite{jura08-1} explored the effects of stellar winds on small asteroids, no  detailed modelling has been done for the stripping of larger objects and it is unclear if the stellar wind during the AGB can provide the level of erosion suggested here and in \cite{melisetal11-1}.

\begin{figure}
  \centering
   \includegraphics[width=7.5cm]{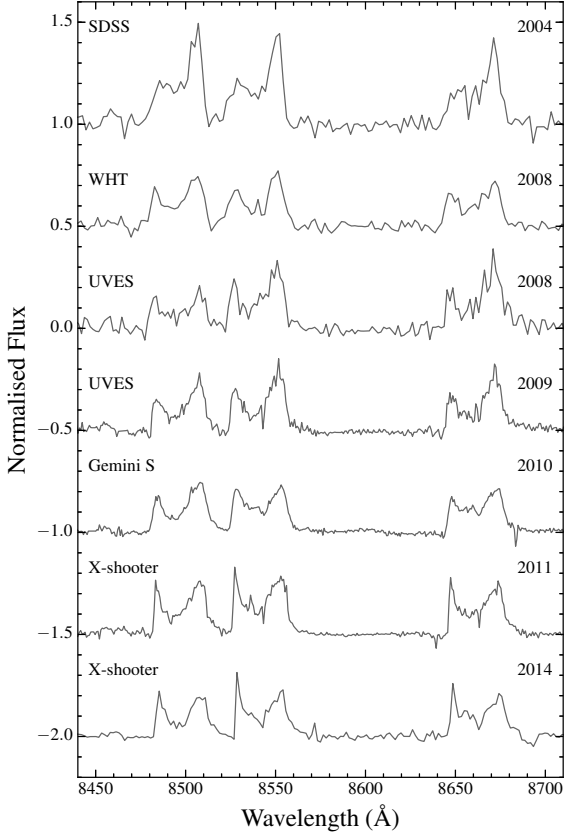}
    \caption{Time-series spectra of the \Ion{Ca}{ii} 8600\,\AA\ emission line triplet, plotted on the same scale with each successive spectrum offset by -0.5. The SDSS data from 2004 show a pronounced asymmetry between the two peaks, with the red peak significantly greater in strength. This difference has since vanished, although in the most recent two spectra the blue peak appears marginally stronger.          
       \protect\label{fig:trips}}
\end{figure}

\begin{figure}
\centering
\includegraphics[width=7.5cm]{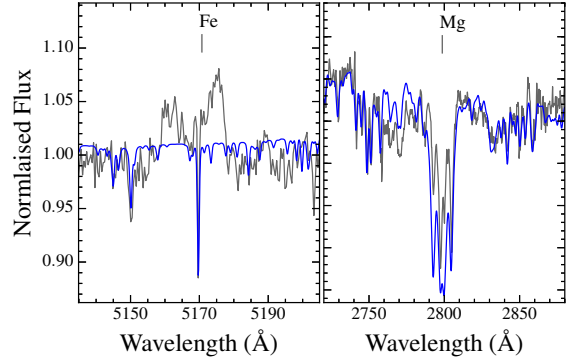}
\caption{Left: Example of weak double-peaked emission from \Ion{Fe}{ii}. Right: The emission from \Ion{Mg}{ii} predicted by \citet{hartmannetal14-1} is not detected, but could be masked by the photospheric absorption lines. The model atmosphere fit (blue) over-predicts the strength of the \Ion{Mg}{ii} absorption, which could be due to some flux contribution by emission from the gaseous disc. \protect\label{fig:gas}}
\end{figure}

\begin{table} 
\centering
\caption{Change in the equivalent width of the \Ion{Ca}{ii} 8600\,\AA\ emission line triplet. The strength of the triplet dropped in the period 2004--2008, but has remained stable to within $1\,\sigma$ since.}
\begin{tabular}{lc}
\hline
Date & \Ion{Ca}{ii} 8600\,\AA\ Equivalent Width (\AA)\\
\hline
2004 December 19 & -16.9$\pm$2.2\\
2008 January 02  & -10.3$\pm$1.3\\
2008 Apri1 03 & -10.8$\pm$1.4\\
2009 April 10 & -11.4$\pm$1.4\\
2010 April 02 & -10.5$\pm$1.3\\
2011 January 29 & -12.8$\pm$1.5\\
2014 October 20 & -8.6$\pm$1.1\\
\hline
\end{tabular}
\label{tab:widths}
\end{table}

\begin{figure*}
    \centering
    \includegraphics[width=17cm]{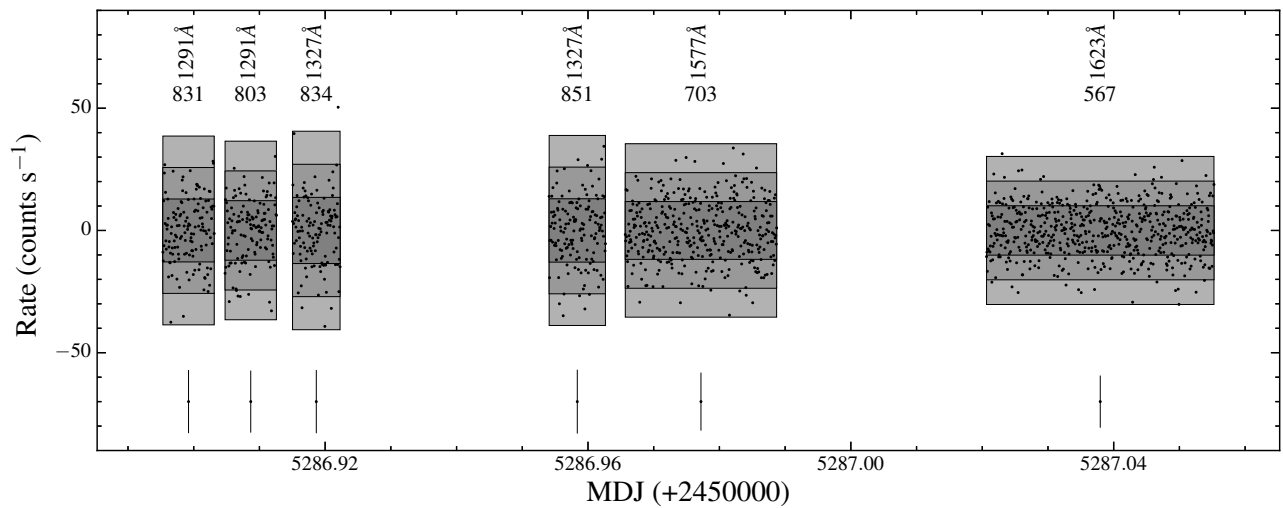}
    \caption{Background-subtracted, normalised light curves of SDSS\,J0845+2257, extracted from the {\em HST}/COS time-tag data.  Dark grey, grey and light grey areas represent 1-$\sigma$ , 2-$\sigma$ and 3-$\sigma$ standard deviations from the mean respectively, assuming a normal distribution for the count rate in each observation. The vertical and horizontal numbers above each spectrum are the central wavelength and mean count rate, respectively. An error bar showing the typical photon noise is shown beneath each spectrum. \protect\label{fig:lc}}
\end{figure*}

\section{Variability of the \Ion{Ca}{ii} Triplet}
\label{sec:gas}
SDSS\,J0845+2257 is part of a rare subset of metal-polluted white dwarfs that show double-peaked emission in the  8600\AA\ \Ion{Ca}{ii} triplet, indicative of a gaseous component to the debris disc \citep{gaensickeetal06-3,gaensickeetal07-1, gaensickeetal08-1,gaensicke11-1, melisetal12-1, farihietal12-1, wilsonetal14-1}.
 The formation and evolution of these gaseous discs is still poorly understood, as the gas resides outside of the sublimation radius, within which gas is naturally expected to be present, and they are found at only a small fraction of the white dwarfs with dusty discs. There is no known case of a purely gaseous disc.  
 
Time-series observations of the 8600\,\AA\ \Ion{Ca}{ii} triplet are shown in Figure \ref{fig:trips}. The double-peaked morphology arises from the Doppler shifts
induced by the Keplerian velocity of the material in the disc
\citep{horne+marsh86-1}, a consequence of which is that the inner and outer
radii of the disc can be estimated from the total width and peak separation
of the \Ion{Ca}{ii} lines respectively. We find $R_\mathrm{in}\sin^2
i\approx0.3\,\Rsun$ and $R_\mathrm{out}\sin^2 i\approx0.8\,\Rsun$,
(where $i$ is the unknown inclination of the disc), consistent with the measurements of \cite{gaensickeetal08-1}. Modelling the observed infrared excess, \cite{brinkworthetal12-1} estimated the extent of the dusty disc to extend from $R_\mathrm{in}\approx0.17\,\Rsun$ to  $R_\mathrm{out}\approx99.7\,\Rsun$, although the outer radius of the disc is unconstrained due to the lack of observations at longer wavelengths. The overlap of these values strongly suggests that the gaseous and dusty components are co-orbital. \cite{hartmannetal14-1} attempted to better constrain the parameters of the gaseous disc component using a non-LTE spectral model, but were unable to obtain a satisfactory fit to the emission lines. 

The debris discs have provided some of the clearest evidence for variability in evolved planetary systems, in particular the rapid appearance and loss of the gaseous disc around SDSS\,J1617+1620 \citep{wilsonetal14-1} and the drop in infrared luminosity from the dust at WD\,J0959--0200 \citep{xuetal14-2}. Similar, although less pronounced, changes are observed in the 8600\,\AA\ \Ion{Ca}{ii} emission line triplet of SDSS\,J0845+2257. Table \ref{tab:widths} lists the change in equivalent widths of the \Ion{Ca}{ii} emission lines over time. The equivalent width was calculated over all three double peaked lines to avoid systematic uncertainties caused by the overlap between the 8495\,\AA\ and 8542\,\AA\ components. The main source of uncertainty in the measurement is the polynomial fit used to normalise the spectrum. To account for this the fit was varied slightly and the resulting scatter in the equivalent width measurements incorporated into the stated error. The data demonstrate a ${\approx}1/3$ drop in the equivalent width of the emission lines between 2004 and 2008, but the strength of the lines has remained  constant to within $1\,\sigma$ since then. In addition, the time-series spectra (Figure \ref{fig:trips}) show pronounced changes in the morphology of the lines. In the 2004 SDSS observation in the red-shifted peaks were substantially stronger, but in more recent observations they have decreased in strength. Conversely the blue-shifted peaks have grown, to the extent that they have become slightly stronger than the red peaks in the latest observations. Due to the relatively low cadence of the time-series observations compared with SDSS\,J1617+1620 \citep{wilsonetal14-1} we can only speculate about the cause of this variability, e.g the interaction with a debris stream formed by a tidally disrupted planetesimal \citep[see for example][]{debesetal12-1,verasetal14-1}.

In addition to the \Ion{Ca}{ii} emission line triplet, some of our spectra show double-peaked emission around the \Ion{Fe}{ii} 5170\AA\ line (Figure \ref{fig:gas}, left panel). The \Ion{Fe}{ii} emission is weaker and less variable than the emission from \Ion{Ca}{ii}, with an average equivalent width of -0.8$\pm$0.5\,\AA. \cite{hartmannetal11-1} predicted  emission from the gaseous disc from 2797\AA\ \Ion{Mg}{ii}, and detected it in the {\em HST} spectrum of SDSS\,J1228+1040. We do not detect emission from \Ion{Mg}{ii} in SDSS\,J0845+2257 (Figure \ref{fig:gas}), although the observed \Ion{Mg}{ii} absorption line triplet is weaker than predicted by the model atmosphere fit, possibly due to a small flux contribution from a disc emission line.  

It is intriguing that the three white dwarfs with the highest and third highest inferred accretion rates, SDSS\,J0738+1835 \citep{dufouretal12-1} and SDSS\,J0845+2257 respectively, both host debris discs with gaseous components. Unfortunately, measurements of the total metal accretion rate have only been made for five gas-hosting white dwarfs to date\footnote{SDSS\,J0738+1835 \citep{dufouretal12-1}, SDSS\,J1228+1040 \citep{gaensickeetal12-1}, He\,1349-2305 \citep{melisetal12-1}, SDSS\,J1617+1620 \citep{wilsonetal14-1} and SDSS\,J0845+2257 (this work).}, so the sample size is too small to reliably test any correlation between the presence of gas and the accretion rate.

\section{Short Term Variability}
\label{sec:light}
We used the time-tag {\em HST}/COS photon event files to construct a light curve. To perform the background subtraction for each of the six individual G130M/G160M observations we defined two regions on the COS detector, one below and one above the spectrum. The 1200\,\AA\ \Ion{Ni}{i}, 1216\,\AA\ Lyman $\alpha$ and 1302\,\AA\ \Ion{O}{i} airglow emission lines were masked and the edges of the detector segments excluded to reduce the instrumental noise. Figure \ref{fig:lc} shows a background-subtracted, normalised light curve binned to 5 seconds. 

To probe for any variability in the flux from SDSS\,J0845+2257 caused by ongoing metal accretion we used a $\chi^{2}$ test, constructing a light curve using a bin size of 32\,ms, i.e. the intrinsic time resolution of COS in time-tag mode. The box for the extraction of the spectrum was defined so that all counts of the target were included, while minimising the amount of background contribution. We set our null hypothesis that each light curve chunk is constant at the mean value.
We find that a constant light curve has an 8--90\,percent probability of having the observed distribution, meaning that we can not reject the hypothesis that the light curve is constant. The large range in probability is caused by the different total exposure times of the individual light curve chunks.

To confirm this result and investigate any contribution to the variability from the background we repeated the process, this time with a wider box for extracting the spectrum to incorporate more background counts, as well as an identical test across the whole light curve. 
The difference in the results is not significant, so we conclude that the {\em HST}/COS data does not show significant variability on time-scales of 32\,ms to $\simeq30$\,minutes.

As is obvious from Figure \ref{fig:lc}, no transits by planets or planetesimals are detected. This non-detection is unsurprising, due to both the short duration of the observations and the fact that close-in planets around white dwarfs are probably rare \citep{veras+gaensicke15-1}.      

\section{Conclusion}
\label{sec:end}
We have carried out a detailed spectroscopic study, over a wide wavelength range, of the metal-polluted white dwarf SDSS\,J0845+2257. The star is accreting debris at a rate of  $1.6\times10^{10}\,\mathrm{g\,s^{-1}}$  and the mass of metals in the convection zone implies a parent body $\gtrsim100$\,km in diameter. Measurements of ten metals have shown that the disrupted planetesimal was similar to the Earth in composition, with a differentiated chemistry dominated by O, Mg, Si and Fe. The relativity high levels of Fe and Ni suggests that the planetesimal may have had a portion of its mantle stripped during the AGB phase, leaving a composition dominated by core material. An unusually large amount of carbon is present, although this could be primordial to the white dwarf. The white dwarf is also orbited by a debris disc with a mildly variable gaseous component.

\section{Acknowledgements}
The authors thank M. Jura and the referee, P. Dufour, for constructive discussions regarding the differences in our studies of this object, and C. Manser for performing the telluric correction to the X-shooter spectrum.  The research leading to these results has received funding from the
European Research Council under the European Union's Seventh Framework
Programme (FP/2007-2013) / ERC Grant Agreement n. 320964 (WDTracer).
 SGP acknowledges financial support from FONDECYT in the form of grant number 3140585. 

This paper is based on observations made with the NASA/ESA {\em Hubble Space Telescope}, obtained at the Space Telescope Science Institute, which is operated by the Association of Universities for Research in Astronomy, Inc., under NASA contract NAS 5-26555. These observations are associated with program ID 11561.

This paper has also made use of observations from the SDSS-III, funding for which has been provided by the Alfred P. Sloan Foundation, the Participating Institutions, the National Science Foundation, and the U.S. Department of Energy Office of Science. 

Observations were also made with the William Herschel Telescope on the island of La Palma by the Isaac Newton Group in the Spanish Observatorio del Roque de los Muchachos of the Instituto de Astrof\'isica de Canarias, the Gemini South Telescope under program GN-2010A-Q-17, and with the ESO VLT at
the La Silla Paranal Observatory under programme IDs  081.C-0466(A), 082.C-0495(A),	383.C-0695(A), 386.C-0218(B) and 094.D-0344(A).

\bibliographystyle{mn_new.bst}
\bibliography{aamnem99,aabib}

\bsp

\end{document}